RESEARCH ARTICLE                    OPEN ACCESS

# A New Approach of Learning Hierarchy Construction Based on Fuzzy Logic

Ali AAJLI*,  Karim AFDEL**
* Laboratory of Computer Systems and Vision -LabSIV, Ibn Zohr University Agadir, Morocco
** Laboratory of Computer Systems and Vision -LabSIV, Ibn Zohr University Agadir, Morocco

**Abstract-**
In recent years, adaptive learning systems rely increasingly on learning hierarchy to customize the educational logic developed in their courses. Most approaches do not consider that the relationships of prerequisites between the skills are fuzzy relationships. In this article, we describe a new approach of a practical application of fuzzy logic techniques to the construction of learning hierarchies. For this, we use a learning hierarchy predefined by one or more experts of a specific field. However, the relationships of prerequisites between the skills in the learning hierarchy are not definitive and they are fuzzy relationships. Indeed, we measure relevance degree of all relationships existing in this learning hierarchy and we try to answer to the following question: Is the relationships of prerequisites predefined in initial learning hierarchy are correctly established or not?

## I. INTRODUCTION

In 1968 Gagne defined the construction of learning hierarchies for programmed instruction (Gagne, 1968; Skinner, 1986; Molenda, 2008) purposes, and in particular, for Branching or Intrinsic Programming (Crowder, 1962; Roe, 1962; Molenda, 2008) which is directly related to a particular view of cognition and learning called behaviorism (Ertmer & Newby, 1993; Greeno, Collins & Resnick, 1996).

Robert Gagne (1968) defined a learning hierarchy as a set of specified intellectual capabilities or intellectual skills. The capabilities in the hierarchy have an ordered relationship to each other and the hierarchy, as a whole, bears some relation to a plan for effective instruction. The hierarchy is built in a manner to reflect that a lower level skill must be acquired or mastered before an upper-level one, that is, lower level capabilities are prerequisites for upper level ones. Intellectual capabilities or skills are the nodes of the hierarchy.  Gagne (1968) defines them as cognitive strategies that denote capabilities for action. Additionally, they also depict a learning route, a path, from simple skills to a final complex capability.

Learning hierarchies not only serve to represent effective instruction plans in terms of skills or capabilities, but also, they serve as diagnosis instruments for providing individual or personalized remediation to students. However, for classrooms with a large number of students, the application of learning hierarchies for individualized (remedial) instruction is a highly time consuming task. Learning hierarchies belong to the behaviorist view on cognition and learning (Ertmer & Newby, 1993; Greeno, Collins & Resnick, 1996), which is a perspective that had, as goals, to make the teaching-learning process more effective and customized to individual differences, in order to improve students' performance on test situations (Molenda, 2008).

The following section presents an overview of some existing approaches for learning hierarchy and discusses their limits.

## II. OVERVIEW OF SOME EXISTING APPROACHES FOR LEARNING HIERARCHY

### 1. Approach by programmed instruction

One approach to apply learning hierarchy in real educational settings is to arrange the content in small steps, or frames of information. These steps lead the learner from the simple to the complex in a carefully ordered sequence, and, most important, at each step the learner is required to make a response, that is, to write or select an answer. This is called programmed instruction (Skinner, 1986; Molenda, 2008) and in its simplest form, which is called linear programming, it represents a linear graph formed by a set of frames, where every frame to the left is a prerequisite for the frames on the right.

However, this view to programmed instruction had and important flaw: all students, regardless of their aptitudes or their prior knowledge of the subject matter, had to go through the same frames and no remedial steps where included.

### 2. Approach by Branching Programming

The development of Branching or Intrinsic Programming is a technique allowed learners to skip ahead through material that was easy for them or to branch off to remedial frames when they had difficulty (Crowder, 1962; Roe, 1962; Molenda, 2008). The ultimate goal of branching programming is to take care of the individual differences of students, in terms of prior knowledge of the subject matter and other abilities that the learner brings (Roe, 1962), and provide personalized paths of learning.

It has to be stated that just as with linear programming, the frames in branching programming, including the remedial ones, had to be designed a priori. This proved to be a very difficult task and led to the design of very complex branching programming graphs and procedures such as: backward branching to missed items, backward branching to review an entire sequence of items, backward branching to alternate form items, lateral branching to supplemental or prerequisite material, lateral branching to supplemental practice items, branching down to a lower level or more detailed items for slow students, branching up to a faster program for bright students, and finally, forward branching by skipping items (Roe, 1962). The complexity of the graphs makes this approach very difficult its practical application.

### 3. Approach by Fuzzy logic

Several learning systems build their learning hierarchies by using a number of different methods of fuzzy logic (Al-Sarem et al, 2010 and Chu et al., 2010 and Chen and Bai, 2008). Sue et al., 2010, used a two-phase method that extracts the association rules between the skills by applying fuzzy logic to convert the grades learners into three levels of difficulty and construct a learning hierarchy. Bai and Chen, 2010, simplified and improved the latter method in adaptive way.

These methods considered grades obtained by learners during the process learning is a fuzzy notion. However, they don't take into account the possibility of using a learning hierarchy predefined by one or more experts of a specific field.

Before introducing our approach, the following section describes some concepts of fuzzy logic which we use later in this paper.

### III. FUZZY SETS THEORY (FST)

Since 1965, the Fuzzy Sets Theory has advanced in a variety of ways and in many disciplines. Fuzzy sets were introduced by Zadeh to represent mathematically the vagueness on certain classes of objects and provide the basis for fuzzy logic.

The fuzzy sets were introduced to model human knowledge representation, and thus improve the performance of systems that use this modelling decision. Fuzzy sets admit gradation such as all tones between black and white. A fuzzy set has a graphical description that expresses how the transition from one to another takes place. This graphical description is called a membership function.

A fuzzy part (or fuzzy set) of a set E is an application $\mu_A(x): E \rightarrow [0, 1]$:

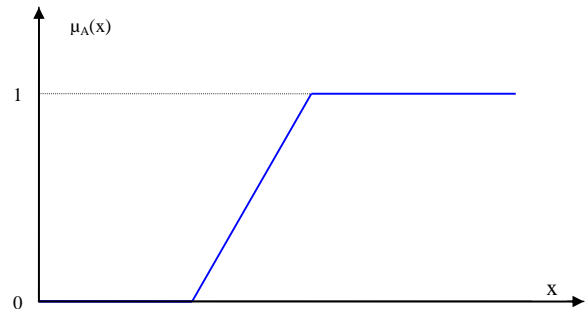

Fig. 1.  $\mu_A(x)$ : A membership function

### IV. OUR PROPOSED APPROACH

In our approach, we describe a new approach of a practical application of fuzzy logic techniques to the construction of Learning Hierarchies.

For this, we use a learning hierarchy predefined by one or more experts of a specific field.

However, the relationships of prerequisites between the skills in the learning hierarchy are not definitive and they are fuzzy relationships.

Indeed, we try -with using fuzzy logic- to answer to the following question: Is the relationships of prerequisites predefined in initial learning hierarchy are correctly established or not?

To respond to this question, we follow the following phases: The first phase determines an initial predefined learning hierarchy, the second phase measure the variation of grades of learners, the next phase transformed the data by using the fuzzification technique, then the next phase mine the association rules between the skills. In the last two phases we propose to build the final learning hierarchy.

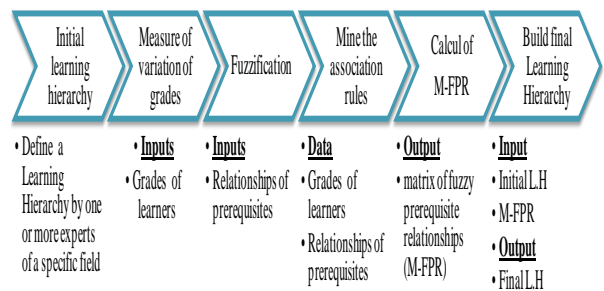

Fig. 2.  Phases of our approach

### 1. Define an initial learning hierarchy

For an expert in a particular field, the presentation of the methodology and sequence to be used for the construction of learning hierarchy is achievable by following the steps below described by Gagne (1968):
Defined a learning hierarchy as a set of specified intellectual capabilities or intellectual skills.
The capabilities in the hierarchy have an ordered relationship to each other and the hierarchy, as a whole, bears some relation to a plan for effective instruction.
The hierarchy is built in a manner to reflect that a lower level skill must be acquired or mastered before an upper-level one, that is, lower level capabilities are prerequisites for upper level ones. Intellectual capabilities or skills are the nodes of the hierarchy.

At the end we will have an initial learning hierarchy as shown in figure below:

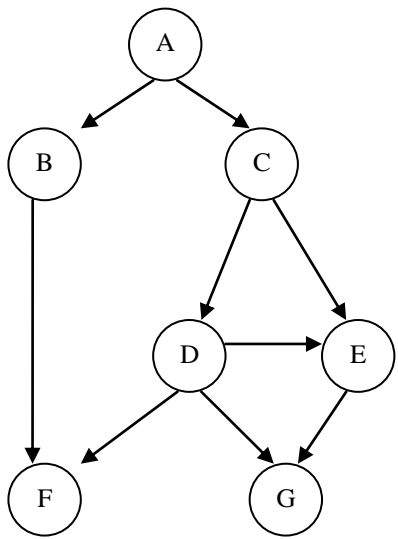

Fig. 3. *Example of an initial learning hierarchy prepared by an expert in a specific field*

The Figure 2 shows an example of a learning hierarchy of a course containing 10 relevant skills, and prerequisite relationships among them.
From the links of the learning hierarchy we define the matrix M of prerequisites between skills, where the value of each element $M_{ij}$ is calculated as below:
$M_{ij} = 1$ means the skill « i » is a prerequisite of the skill « j ».
$M_{ij} = 0$ means the skill « i » is not a prerequisite of the skill « j ».
« i » represents the rows and « j » the columns.

Table 1 below, shows a matrix representation ($M_{ij}$) of initial predefined learning hierarchy of the figure 2.

For example, the first line means that the skill A is a prerequisite of the skills B and C.

TABLE I. MATRIX REPRESENTATION OF INITIAL PREDEFINED LEARNING HIERARCHY

| $M_{ij}$ | A | B | C | D | E | F | G |
|---|---|---|---|---|---|---|---|
| A | 0 | 1 | 1 | 0 | 0 | 0 | 0 |
| B | 0 | 0 | 0 | 0 | 0 | 1 | 0 |
| C | 0 | 0 | 0 | 1 | 1 | 0 | 0 |
| D | 0 | 0 | 0 | 0 | 1 | 1 | 1 |
| E | 0 | 0 | 0 | 0 | 0 | 0 | 1 |
| F | 0 | 0 | 0 | 0 | 0 | 0 | 0 |
| G | 0 | 0 | 0 | 0 | 0 | 0 | 0 |

## 2. Variation of grades
### A. Retrieving digital data

In this sub-phase, we retrieve the numerical grades obtained during assessments of each student in each skill in a learning process. These grades are collected in a matrix called the matrix grades: Grades (Learner (Si), Skill (i)) such as:

TABLE II. EXAMPLE OF MATRIX GRADES OF 10 STUDENTS

| Grades | A | B | C | D | E | F | G |
|---|---|---|---|---|---|---|---|
| $S_1$ | 10 | 10 | 1 | 3 | 7 | 9 | 3 |
| $S_2$ | 11 | 12 | 5 | 7 | 11 | 11 | 7 |
| $S_3$ | 10 | 11 | 5 | 3 | 8 | 10 | 5 |
| $S_4$ | 13 | 10 | 6 | 6 | 10 | 10 | 10 |
| $S_5$ | 15 | 18 | 10 | 12 | 16 | 16 | 15 |
| $S_6$ | 19 | 18 | 6 | 10 | 14 | 19 | 13 |
| $S_7$ | 12 | 11 | 1 | 5 | 6 | 10 | 4 |
| $S_8$ | 3 | 4 | 0 | 2 | 5 | 7 | 5 |
| $S_9$ | 15 | 16 | 6 | 10 | 11 | 18 | 13 |
| $S_{10}$ | 12 | 14 | 5 | 3 | 0 | 13 | 0 |

Table 2 shows an example of 10 students and their grades within 7 skills that constitute initial learning hierarchy.
Where:
The maximum score that a student can have in an assessment is equal to 20.

### B. Measure of variation of grades
In this sub-phase, we measure the variation of grades of all prerequisite relationships of initial predefined learning hierarchy.
The Matrix of variation of grades ΔGrades (i, j) is calculated using the both matrix:
- Matrix Grades $_{(Learner (Si), Skill (i))}$
- Matrix Mij

ΔGrades (i, j) $_{Learner}$ = [Grade (j) – Grade (i)] with $M_{ij}$ = 1 i.e the skill « i » is a prerequisite of the skill « j ».

And $-20 \leq \Delta Grades \leq 20$

In table bellow we proposer an example of matrix ΔGrades (i, j) based on the data of the tables 1 and 2:

TABLE III.    MATRIX OF VARIATION OF GRAGES OF INITIAL MAP (ΔGRADES)

| ΔGrades | A↓B | A↓C | B↓F | C↓D | C↓E | D↓E | E↓G | D↓G | D↓F |
|---|---|---|---|---|---|---|---|---|---|
| $S_1$ | 0 | -9 | -1 | 2 | 6 | 4 | -4 | 0 | 6 |
| $S_2$ | 1 | -6 | -1 | 2 | 6 | 4 | -4 | 0 | 4 |
| $S_3$ | 1 | -5 | -1 | -2 | 3 | 5 | -3 | 2 | 7 |
| $S_4$ | -3 | -7 | 0 | 0 | 4 | 4 | 0 | 4 | 4 |
| $S_5$ | 3 | -5 | -2 | 2 | 6 | 4 | -1 | 3 | 4 |
| $S_6$ | -1 | -13 | 1 | 4 | 8 | 4 | -1 | 3 | 9 |
| $S_7$ | -1 | -11 | -1 | 4 | 5 | 1 | -2 | -1 | 5 |
| $S_8$ | 1 | -3 | 3 | 2 | 5 | 3 | 0 | 3 | 5 |
| $S_9$ | 1 | -9 | 2 | 4 | 5 | 1 | 2 | 3 | 8 |
| $S_{10}$ | 2 | -7 | -1 | -2 | -5 | -3 | 0 | -3 | 10 |

### 3. Prerequisite relationships fuzzification

The fuzzy set theory is used to simplify the analysis of the numerical results of the evaluations of learners with transforming their digital data in membership functions.

In our approach this theory is applied to the prerequisite relationships of initial learning hierarchy.

Let X a set of prerequisite relationships of initial learning hierarchy.
Let CPR a fuzzy subset of prerequisite relationships that can be classified as a correct prerequisite relationships between skill « i » and skill « j ».

$$CPR = \{(k, \mu_{CPR}(k))/k \in X\}$$

Where:

$\mu_{CPR}(k)$ Is the membership function of CPR, the values of this function present the relevance degree of each link « k » in the fuzzy set CPR.

Let RPR a fuzzy subset of links that can be classified as wrong prerequisite relationships between skill « i » and skill « j », but can be classified also as a correct prerequisite relationships between skill « j » and skill « i ».

$$RPR = \{(k, \mu_{RPR}(k))/k \in X\}$$

Where:

$\mu_{RPR}(k)$ is the membership function of RPR, the values of this function present the relevance degree of each link « k » in the fuzzy set RPR.

The definition of the two membership functions of fuzzy sets $\mu_{CPR}(k)$ and $\mu_{RPR}(k)$ is based on the indicator expressed as « variation of grades of all prerequisite relationships of initial predefined learning hierarchy (ΔGrades) » (this indicator is calculated in the above section "*Measure of variation of grades*").

### 4. Mine the association rules between the skills

For mining the association rules between the skills we use the following table:

| Rule | Prerequisite relationships (k) |
|---|---|
| S1 ≤ ΔGrades ≤ S2 {S1 < 0, S2>0} | $k \in CPR$ |
| S2 ≤ ΔGrades ≤ S3 {S3 > S2} | $k \in RPR$ |

Then, the two functions $\mu_{CPR}(k)$ and $\mu_{RPR}(k)$ are based on the above rules and they are defined as below:

$$\mu_{CPR}(k) = \begin{cases} 0 & \text{if } \Delta Grades < S1 \\ \frac{-1}{S1}\Delta Grades + 1 & \text{if } S1 \leq \Delta Grades \leq 0 \\ \frac{-1}{S2}\Delta Grades + 1 & \text{if } 0 < \Delta grades \leq S2 \\ 0 & \text{if } \Delta Grades > S2 \end{cases}$$

$$\mu_{RPR}(k) = \begin{cases} 0 & \text{if } \Delta Notes < 0 \\ \frac{1}{S2}\Delta Notes & \text{if } 0 \leq \Delta Notes \leq S2 \\ \frac{-(\Delta Notes + S3)}{S3 - S2} & \text{if } S2 < \Delta Notes \leq S3 \\ 0 & \text{if } \Delta Notes > S3 \end{cases}$$

Where:
The three thresholds S1, S2 and S3 are defined in collaboration with experts in the field studied.
Based on our experience feedback the threshold values are chosen as follows:
S1 = variation of -5 grades
S2 = variation of 5  grades
S3 = variation of 10 grades
Then the two functions $\mu_{CPR}(k)$ and $\mu_{RPR}(k)$ becomes:

$$\mu_{CPR}(k) = \begin{cases} 0 & \text{if } \Delta Grades < -5 \\ \frac{1}{5}\Delta Grades + 1 & \text{if } -5 \leq \Delta Grades \leq 0 \\ \frac{-1}{5}\Delta Grades + 1 & \text{if } 0 < \Delta grades \leq 5 \\ 0 & \text{if } \Delta Grades > 5 \end{cases}$$

$$\mu_{RPR}(k) = \begin{cases} 0 & \text{if } \Delta Grades < 0 \\ \frac{1}{5}\Delta Grades & \text{if } 0 \leq \Delta Grades \leq 5 \\ \frac{-1}{5}\Delta Grades + 2 & \text{if } 5 < \Delta grades \leq 10 \\ 0 & \text{if } \Delta Grades > 10 \end{cases}$$

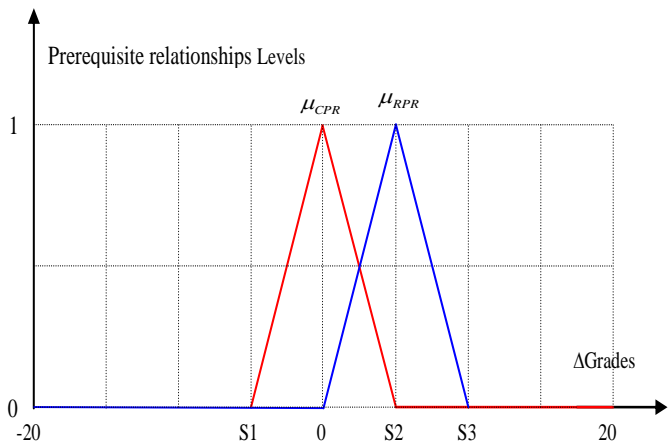

Fig. 4. membership functions

### 5. Results of prerequisite relationships fuzzification

Table 4 shows the result of prerequisite relationships fuzzification.

This result will be denoted matrix of fuzzy prerequisite relationships (M-FPR).

TABLE IV. RESULT OF PREREQUISITE RELATIONSHIPS FUZZIFICATION

| | A↓B | | A↓C | | B↓F | |
|---|---|---|---|---|---|---|
| | μ(CPR) | μ(RPR) | μ(CPR) | μ(RPR) | μ(CPR) | μ(RPR) |
| $S_1$ | 1,00 | 0,00 | 0,00 | 0,00 | 0,80 | 0,00 |
| $S_2$ | 0,80 | 0,20 | 0,00 | 0,00 | 0,80 | 0,00 |
| $S_3$ | 0,80 | 0,20 | 0,00 | 0,00 | 0,80 | 0,00 |
| $S_4$ | 0,40 | 0,00 | 0,00 | 0,00 | 1,00 | 0,00 |
| $S_5$ | 0,40 | 0,60 | 0,00 | 0,00 | 0,60 | 0,00 |
| $S_6$ | 0,80 | 0,00 | 0,00 | 0,00 | 0,80 | 0,20 |
| $S_7$ | 0,80 | 0,00 | 0,00 | 0,00 | 0,80 | 0,00 |
| $S_8$ | 0,80 | 0,20 | 0,40 | 0,00 | 0,40 | 0,60 |
| $S_9$ | 0,80 | 0,20 | 0,00 | 0,00 | 0,60 | 0,40 |
| $S_{10}$ | 0,60 | 0,40 | 0,00 | 0,00 | 0,80 | 0,00 |
| **AVG** | **0,72** | **0,18** | **0,04** | **0,00** | **0,74** | **0,12** |

| | C↓D | | C↓E | | D↓E | |
|---|---|---|---|---|---|---|
| | μ(CPR) | μ(RPR) | μ(CPR) | μ(RPR) | μ(CPR) | μ(RPR) |
| $S_1$ | 0,60 | 0,40 | 0,00 | 0,80 | 0,20 | 0,80 |
| $S_2$ | 0,60 | 0,40 | 0,00 | 0,80 | 0,20 | 0,80 |
| $S_3$ | 0,60 | 0,00 | 0,40 | 0,60 | 0,00 | 1,00 |
| $S_4$ | 1,00 | 0,00 | 0,20 | 0,80 | 0,20 | 0,80 |
| $S_5$ | 0,60 | 0,40 | 0,00 | 0,80 | 0,20 | 0,80 |
| $S_6$ | 0,20 | 0,80 | 0,00 | 0,40 | 0,20 | 0,80 |
| $S_7$ | 0,20 | 0,80 | 0,00 | 1,00 | 0,80 | 0,20 |
| $S_8$ | 0,60 | 0,40 | 0,00 | 1,00 | 0,40 | 0,60 |
| $S_9$ | 0,20 | 0,80 | 0,00 | 1,00 | 0,80 | 0,20 |
| $S_{10}$ | 0,60 | 0,00 | 0,00 | 0,00 | 0,40 | 0,00 |
| **AVG** | **0,52** | **0,40** | **0,06** | **0,72** | **0,34** | **0,60** |

| | E↓G | | D↓G | | D↓F | |
|---|---|---|---|---|---|---|
| | μ(CPR) | μ(RPR) | μ(CPR) | μ(RPR) | μ(CPR) | μ(RPR) |
| $S_1$ | 0,20 | 0,00 | 1,00 | 0,00 | 0,00 | 0,80 |
| $S_2$ | 0,20 | 0,00 | 1,00 | 0,00 | 0,20 | 0,80 |
| $S_3$ | 0,40 | 0,00 | 0,60 | 0,40 | 0,00 | 0,60 |
| $S_4$ | 1,00 | 0,00 | 0,20 | 0,80 | 0,20 | 0,80 |
| $S_5$ | 0,80 | 0,00 | 0,40 | 0,60 | 0,20 | 0,80 |
| $S_6$ | 0,80 | 0,00 | 0,40 | 0,60 | 0,00 | 0,20 |
| $S_7$ | 0,60 | 0,00 | 0,80 | 0,00 | 0,00 | 1,00 |
| $S_8$ | 1,00 | 0,00 | 0,40 | 0,60 | 0,00 | 1,00 |
| $S_9$ | 0,60 | 0,00 | 0,40 | 0,60 | 0,00 | 0,40 |
| $S_{10}$ | 1,00 | 0,00 | 0,40 | 0,00 | 0,00 | 0,00 |
| **AVG** | **0,66** | **0,00** | **0,56** | **0,36** | **0,06** | **0,64** |

### 6. Build the final learning hierarchy
   A.   *Algorithm*

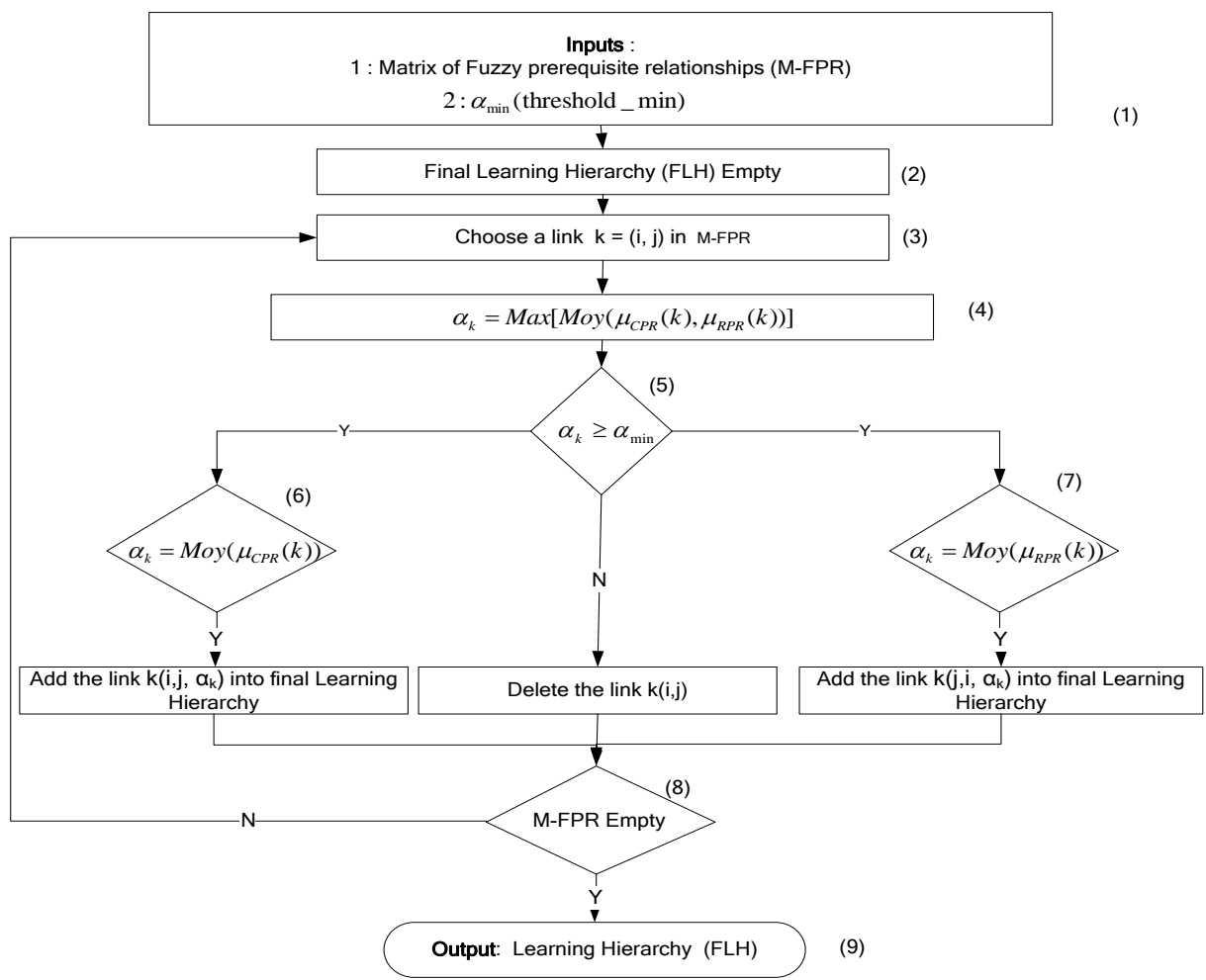

Fig. 5.   Algorithm of learning hierarchy constructing process

### B. Final learning hierarchy

In last step we use the algorithm above for mining the prerequisite relationships with their relevance degree and generate the final learning hierarchy.
Input data of the algorithm are:
- Matrix of fuzzy prerequisite relationships (M-FPR)
- A threshold minimum of prerequisite relationships is a threshold that indicates the prerequisite relationships meaningful in the construction process.

At first, the final learning hierarchy is empty.
For each link « k » existing in the matrix of fuzzy prerequisite relationships we test:

If the value of maximum of average of each membership functions $\mu_{CPR}(k)$ and $\mu_{RPR}(k)$ is greater or not than the threshold minimum.
At the end, the link (k) may be:

- Add in the final learning hierarchy in the same direction between his two skills with a relevance degree equal to $\alpha_k$.
- Add in the final learning hierarchy in the opposite direction of the initial link with a relevance degree equal to $\alpha_k$.
- Delete and it is not included in the final learning hierarchy.

### 7. Example of learning hierarchy constructing process

We apply this algorithm to the data (M-FPR) of the table 4
Input data of the algorithm are:
- Matrix of fuzzy prerequisite relationships (M-FPR) of table 4.
- A threshold minimum $\alpha_k=0,5$

Thus, the final learning hierarchy is:

| Initial L.H | A<br>↓<br>B | | A<br>↓<br>C | | B<br>↓<br>F | |
|---|---|---|---|---|---|---|
| | μ(CPR) | μ(RPR) | μ(CPR) | μ(RPR) | μ(CPR) | μ(RPR) |
| AVG | 0,72 | 0,18 | 0,04 | 0,00 | 0,74 | 0,12 |
| Degree of relevance | 0,72 | | - | | 0,74 | |
| Relationships | kept link | | deleted link | | kept link | |
| Final L.H | A<br>↓<br>B | | - | | B<br>↓<br>F | |

| Initial L.H | C<br>↓<br>D | | C<br>↓<br>E | | D<br>↓<br>E | |
|---|---|---|---|---|---|---|
| | μ(CPR) | μ(RPR) | μ(CPR) | μ(RPR) | μ(CPR) | μ(RPR) |
| AVG | 0,52 | 0,40 | 0,06 | 0,72 | 0,34 | 0,60 |
| Degree of relevance | 0,52 | | 0,72 | | 0,60 | |
| Relationships | kept link | | substituted link | | substituted link | |
| Final L.H | C<br>↓<br>D | | E<br>↓<br>C | | E<br>↓<br>D | |

| Initial L.H | E<br>↓<br>G | | D<br>↓<br>G | | D<br>↓<br>F | |
|---|---|---|---|---|---|---|
| | μ(CPR) | μ(RPR) | μ(CPR) | μ(RPR) | μ(CPR) | μ(RPR) |
| AVG | 0,66 | 0,00 | 0,56 | 0,36 | 0,06 | 0,64 |
| Degree of relevance | 0,66 | | 0,56 | | 0,64 | |
| Relationships | kept link | | kept link | | substituted link | |
| Final L.H | E<br>↓<br>G | | D<br>↓<br>G | | F<br>↓<br>D | |

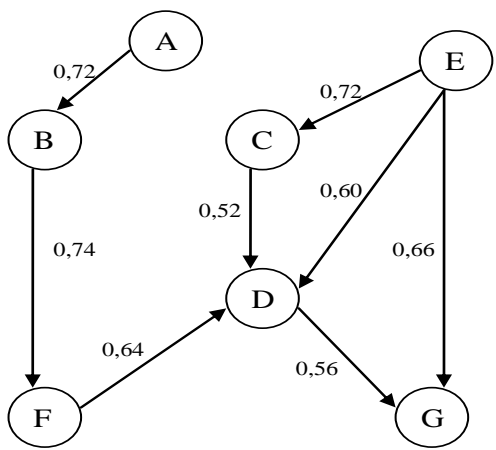

Fig. 6. Final learning hierarchy

## V. CASE STUDY

In this section, we propose an implementation of our approach in Java programming language field.

### 1. Skills chosen for the course of the JAVA programming language

For this course were selected following 12 skills:
1) Elementary of Java
2) Objects and Classes
3) Packages
4) Inner Classes
5) Flux I/O
6) Exceptions
7) Inheritance
8) Serialization
9) Interfaces
10) Polymorphism
11) Threads
12) Collections

### 2. Initial learning hierarchy of the JAVA programming language

Figure below shows the initial learning hierarchy selected:

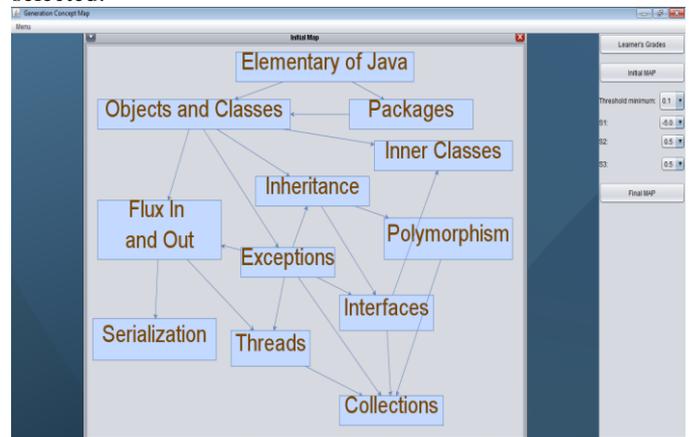

Fig. 7. Initial learning hierarchy of Java

### 3. Generating the final learning hierarchy of JAVA programming language

For this case study we have chosen a minimum $\alpha_k=0,5$

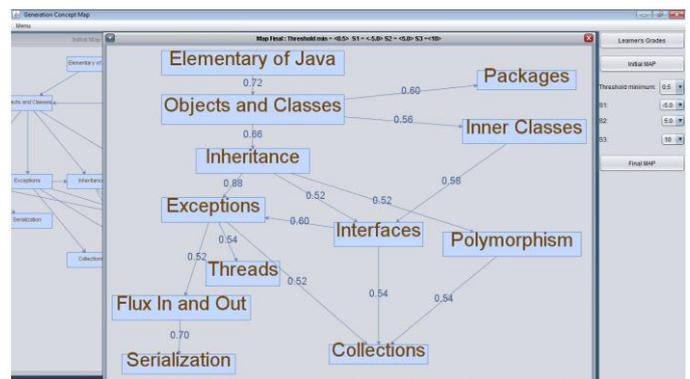

Fig. 8. Final learning hierarchy of Java

## VI. CONCLUSION

In this paper we present a new hybrid approach to construct the learning hierarchy of a specific field, this approach is based on using a predefined expert learning hierarchy and we measure the degree of relevance of all relationships existing in this predefined expert learning hierarchy. This new approach improves the educational protocol to obtain two kinds of prerequisite relationships, the first type can be classified as relationships correctly established by the expert. These relationships must be kept in the final learning hierarchy. The second type can be considered as relations incorrectly established by the expert, these relations must be deleted or substituted with the inverse of the original relationships. For the second type we conclude that there is no correlation between the results obtained and the skills of learners, which can be explained by one or both of the following reasons:

- The use of inappropriate items in the tests of the two skills
- The two skills of this relationship are completely independent.

The results obtained from the application of this new approach on the course of JAVA programming language are good.